\documentclass[onecollarge,natbib]{svjour2}
\bibpunct{[}{]}{;}{n}{}{,} 
\smartqed  

\usepackage{graphics}
\usepackage{graphicx}
\usepackage{dcolumn}
\usepackage{float}
\usepackage{color}
 
\newcommand{\ar}{\arrowvert}

\newcommand{\be}{\begin{equation}} 
\newcommand{\ee}{\end{equation}} 
\newcommand{\ba}{\begin{eqnarray}} 
\newcommand{\ea}{\end{eqnarray}}

\journalname{Few Body Systems}

\begin{document}

\title{To what extent is Gluon Confinement \\ an empirical fact?}
\author{R. L. Delgado, Carlos Hidalgo-Duque and Felipe J. Llanes-Estrada} 
\institute{Dept.  F\'{\i}sica Te\'orica I and II, Univ. Complutense de Madrid, 28040 Madrid, Spain.
}

\maketitle

\begin{abstract}
Experimental verifications of Confinement in hadron physics have established the absence of charges with a fraction of the electron's charge by studying the energy deposited in ionization tracks at high energies, and performing Millikan experiments with charged droplets at rest.  These experiments test only the absence of particles with fractional charge in the asymptotic spectrum, and thus ``Quark''  Confinement.\\
However what theory suggests is that {\emph{Color}} is confined, that is, all asymptotic particles are color singlets. Since QCD is a non-Abelian theory, the gluon force carriers (indirectly revealed in hadron jets) are colored.\\
We empirically examine what can be said about Gluon Confinement based on the lack of detection of appropriate events, aiming at an upper bound for high-energy free-gluon production.
\end{abstract} 

\maketitle
\section{Introduction}

As stated in the abstract, both Millikan oil-drop type experiments~\cite{Perl:2009zz} and ionization track measurements~\cite{Nakamura:2010zzi} are devised to test the absence of electric charges that are fractions of the electron's charge. No evidence has been found for free quarks and bounds have been put on their existence. There are at least four works~\cite{expquark1,expquark2,expquark3,expquark4} that constrain the free quark production cross-section in proton collisions at various energies to be less than $10$ fbarn.

Gluon confinement is simultaneously necessary if modern theoretical pictures of confinement make any sense~\cite{nuestrolibro}, as flux tubes are expected to form between color sources and sinks, and cutting them produces new flux tubes, but not isolated such sources (and gluons do carry color in Quantum Chromodynamics, thus acting as sources). In fact, independently of certain details on infrared behavior that are being sorted out, the theoretical community consensus~\cite{Alkofer:2003jj,Cucchieri:2007rg,Fischer:2009tn} via lattice computations, Dyson-Schwinger and Renormalization Group equations, is that the gluon propagator shows violations of reflection positivity, showing the gluon cannot appear at distances beyond about a Fermi without hadronizing.

However there are no bounds known to us that put quantitative limits to gluon production, beyond the vague statement ``no gluons have been seen as free particles''. No such quantitative experimental statements are presently made about the presence or absence of colored, neutral particles.
This is not a totally satisfactory state of experimental affairs, and the purpose of this note is to help experimental collaborations establish a maximum in the possible gluon production cross section based on contemporary experimental data.

In order to do so, we will consider the (unlikely) hypothesis that gluons are being produced in accelerator reactions such as~\footnote{Color conservation requires that $X$, $Y$ carry color also in the adjoint representation, so that an additional gluon, $q\bar{q}$ pair, or some exotic color cluster, is necessary, but we wish to test color confinement and not color conservation, so that we will further ignore the recoiling $X$, $Y$ systems.}
\be \label{production}
\left\{  \begin{array}{ccc}
e^+ + e^-  & \rightarrow & g + X \\
p +  p & \rightarrow & g + Y \\
\end {array} \right.
\ee
 As a working example we will take the ALICE experiment at the LHC~\cite{Aamodt:2011zj}, but our arguments can easily be extended to other collaborations that have recently taken ample data.

We will try to quantify how the information ``no gluon-like event was reported in excess of background'' translates in terms of a bound on the semiinclusive production cross-section $\sigma_{pp\to g+X}$ with the gluon in the asymptotic state.
We will argue that modern experiments can establish bounds on the relevant cross-section of at most
\be \label{result}
\sigma\left(E_g \ge \ 12 {\rm GeV}\right)_{pp\to g+X} \le O(10{\rm fbarn})
\ee
for gluon energies $E_g$ larger than 12 $GeV$ in the laboratory frame and a large range of center of mass energies, although present methods are still far from reaching that level of exclusion.

\section{Setting collider bounds on free gluon production}

\begin{figure}[H]
\begin{center}
\includegraphics[width=0.15\textwidth]{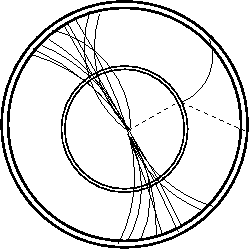}
\end{center}
\caption{{Scheme of a hypothetical gluon (dotted line) escaping the central collision point and producing a secondary collision at the beampipe or Internal Tracking System, where it would materialize with a signature similar to a neutron.\label{fig:esquemita}}}
\end{figure}

Let us therefore follow what would happen to a hypothetical liberated gluon after reaching out beyond a few Fermi from the primary collision point where the two beam protons produced it. The situation is depicted in figure~\ref{fig:esquemita}.
The gluon would propagate in vacuum until reaching, first the beampipe, then the innermost detector (the Internal Tracker in the case of the ALICE setup).
As a neutral particle, the gluon would leave no ionization trace as it would not interact with atomic electrons. But as carrier of the strong nuclear force, it would immediately interact with a nucleon of the medium's material.

Since color is a conserved quantity in Chromodynamics, this secondary interaction would have  to produce either another gluon (perhaps the same, as in elastic scattering) or a quark-antiquark pair in a color octet.
One could therefore probably be able to turn free-quark searches into free-gluon searches if the relevant cross-section was known, but we do not see an immediate way to estimate it. \\
For this reason we rather concentrate on the total gluon-nucleon interaction, for which, as we will argue, there are reasonable theoretical arguments to guess the would be cross-section. 

Nucleons in general would be spallated off an atomic nucleus $A$ in the material; in every sufficiently energetic reaction at least one nucleon is taken away from the proton
and, baryon number being conserved, it propagates.
Since the neutron itself would leave no ionization trace, we will concentrate on gluon-proton interactions, and thus the secondary reaction of choice to implement gluon non-detection is
\begin{equation} \label{main_reaction}
g + A \rightarrow{} g' + p + X
\end{equation}
where $X$ represents the (unmeasured) nuclear remainder, $g'$ should be the undetected, scattered gluon due to color conservation (eventually, $q\bar{q}$ pair) and $p$ is the secondary proton that tags the incoming neutral particle.
The main background in this channel is obviously induced by neutral particles, especially neutrons, and to a much lesser extent 
photons, long-lived kaons and perhaps neutral pions that, as the would be gluon, appear suddenly at the secondary vertex in the material via
\be
n+A \rightarrow{} p + X \ .
\ee

In section~\ref{sec:sigmasec} we will argue, based on Regge theory and color scaling, that the total gluon-proton cross-section is about 3 barn at high energies, much larger than the approximately 50 mbarn of the background nucleon-nucleon cross-section. This means that there is ample room to set limits on gluon production above the background.

One may wonder why we accept the detection reaction in Eq.~(\ref{main_reaction}) to have a non-zero cross-section whereas we assume (and hope) that the production reaction in Eq.~(\ref{result}) will have exactly zero cross-section, since after all both involve asymptotic colored particles. \\
The answer comes from common understanding of color confinement~\cite{Swift:1983fz} that suggests an infinite energy gap between states where all asymptotically separated clusters are color singlets, and those where some are actually colored. 
Whereas in Eq.~(\ref{result}) the reaction proceeds from a color singlet state to a state with distinct color clusters, thus being closed kinematically if our theoretical understanding is correct, in Eq.~(\ref{main_reaction}) both sides are colored, and thus the reaction is kinematically allowed (barring the question of how one prepared the initial state in the first place). Of course we could be surprised by finding out that the reaction in Eq.~(\ref{result}) is actually taking place, against our theoretical prejudice. \\
Another way of casting the hypothesis that we are testing is thus whether a free, deconfined gluon at infinity has or not an infinite mass (this should not be confused with the effective inertia due to interactions while on a bound state).

To proceed let us argue that the gluon mean free-path is very short and thus all putative (high energy) gluons collide early-on in the inner material of the experiment.
Indeed, from elementary theory
\begin{eqnarray}
\lambda_{\rm max}=\frac{1}{n_p ~ \sigma_{\rm min}} \ .
\end{eqnarray}
We estimate $\sigma_{\rm min}\simeq 3.2\ {\rm barn}$ in section~\ref{sec:sigmasec} below, 
and the proton density of Beryllium (commonly used in accelerator beampipes due, precisely, to its low density $\rho =1.85~ \frac{g}{cm^{3}}$ among other features) is 
{
$n_p = 4.9\cdot 10^{23}~ \frac{\rm protons}{cm^{3}}$.
} Thus in beryllium  we obtain 
$ \lambda_{max}=0.6~ cm $.
In a typical silicon composite in the inner tracker, on the other hand, $n=7.0\cdot 10^{23}~ \frac{\rm protons}{cm^{3}}$ and $\lambda = 0.5~ cm$. 

In the particular case of the ALICE experiment, the secondary proton can be identified by the track's energy deposition $dE/dx$, but best identification advises that the proton be transferred a momentum of at least 275 MeV, so it may reach the Time of Flight (TOF) detector in the presence of ALICE's 0.5 Tesla magnetic induction field. Slower protons have a trajectory with curvature radius that does not allow them to reach that detector. Without TOF identification, chances of misidentification increase, and since a charged pion could stem from the decay of neutral $\Lambda$ or other resonances, and we want to avoid increasing the background that needs to be controlled, we would recommend that in the secondary collision the momentum transferred to the proton exceeds those 275 MeV.
{
Thus we will take a subtracted cross section $\sigma_{gp}(s)-\sigma_{gp}(s,t\ge t_0)$ with $t_0=(p_{p'}-p_p)^2= (E'-M_N)^2-{\bf p}^{'2}$, that numerically is $t_0=-(190\ {\rm MeV})^2$ (as obtained from Eq.~(\ref{pomeroneq}) below). This will reduce the secondary cross-section to 2.9 barn which is similar enough to the previous result.
}

So, as the material thicknesses in the innermost workings of the experiment are several centimeters, we can safely assume that every gluon that might be produced collides with a nucleus in the beampipe or inner tracker and ejects a secondary proton or neutron.

The number of gluons that could be reaching the material would be the product of the integrated luminosity of the accelerator during observation time and the deconfining cross-section being tested, namely
\begin{equation}\label{number_primaries}
N_g = L \cdot \sigma_{pp\rightarrow g+X} \ .
\end{equation}

The number of secondary protons emitted  $N_{p}$  by the reaction in Eq.~(\ref{main_reaction}) that we could expect for each incident free gluon
is then readily shown to be
\begin{equation}\label{number_secondaries}
\frac{N_{p}}{N_g} = {\mathcal E_f}(t)  \frac{\Omega}{4\pi} f_{gp}\ .
\end{equation}
In this formula ${\mathcal E_f}(E)$ is the energy-dependent proton detection efficiency, that we estimate to be perfect if the proton momentum exceeds the $275\ MeV$ threshold, but null for less energetic secondaries,
${\mathcal E_f}(E)=\theta(t_0-t)$.\\
Also $\frac{\Omega}{4\pi}$ is the detectors solid-angle acceptance. {For ALICE, that detects particles in the pseudorapidity range $(-1,1)$ or about 46 degrees in polar angle, with full azimuthal coverage, $\Omega/(4\pi)\simeq 0.76$.}\\
The last factor is $0< f_{gp}<1$,  the fraction of collisions where the gluon actually collides with a proton instead of a neutron (as a spallated neutron would be less easily detected, we neglect such interactions). For the light elements such as $Be$ and $Si$ that we are considering, $f_{gp}\simeq \frac{1}{2}$.

If the experimental collaboration could understand the background, and in the face of negative detection of a secondary proton above that background, namely $N_{p}<1$, we would obtain using Eqs.~(\ref{number_primaries}) and~(\ref{number_secondaries}) 
\be
 \label{finalcross}
\sigma_{pp\rightarrow g+X}(s,E_g) \le \frac{4\pi}{L {\mathcal E_f}(t) \Omega  f_{gp}} 
\ee

Which is the wanted bound on the cross-section. This is a function of two arguments, the proton-proton collision energy $s$ and the deconfined gluon's energy $E_g$. All parameters are determined by the experimental setup 
{
except this $E_g$ that we will require to be larger than $12\ GeV$, corresponding to an approximate Mandelstam's $s$ for the secondary reaction
$s_{gp}>(5\ {\rm GeV})^2$  }
to guarantee that we can employ Regge theory to control the secondary reaction's cross section. Our bound can probably be extended well below this cut, but perhaps this is worth a more careful analysis in the future.

The numerical value of the gluon production cross section according to Eq.~(\ref{finalcross}) is $3.5$ fbarn (as reflected in Eq.~(\ref{result}) ) using as integrated luminosity $L \simeq 760~pb^{-1}$ (that the LHC accumulated by June 5th 2011). 

This ultimate bound on free gluon production can be reached in later studies if subsequent channels are analyzed, such as the (not found) free-quark channel. For the time being, our goal of employing tagged protons will allow only for looser bounds.

We also stress that the fraction of gluons lost that do not eject a proton has been accounted for by $f_{gp}$ (that excludes those gluons that collide with a neutron instead) and the momentum transfer of 275 MeV, that providing a minimum of 40 MeV kinetic energy to the proton, makes its 7 MeV binding to a light nucleus irrelevant.

\section{Pythia simulation of neutron background}

To show a possible signal shape and the intensity of the dominant neutron background,
we have prepared a Monte Carlo simulation of a proton-proton collision at 8 TeV in the center of mass. We have adopted a field standard, PYTHIA 8~\cite{sjostrand} without modifications.

We have employed Pythia's options to allow the Monte Carlo to propagate all hard QCD and  soft QCD processes, as well as all heavy quark (c,b,t) production and decay processes, that can give rise to copious hadron production.

\begin{figure}
\begin{center}
\includegraphics[width=0.45\textwidth]{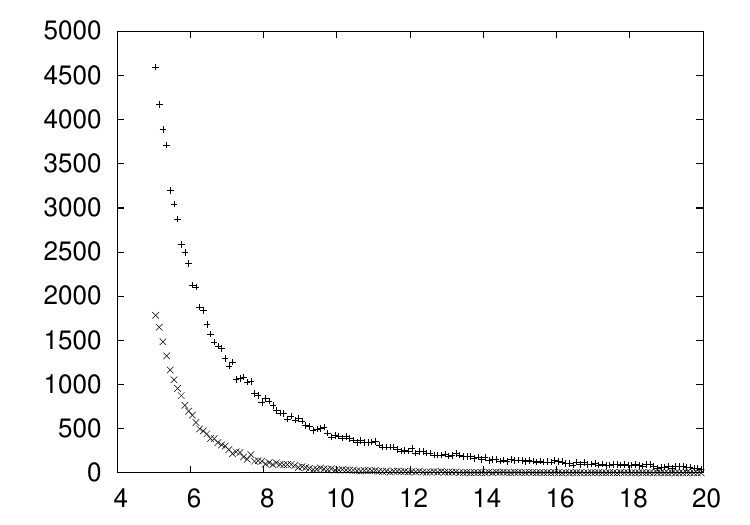}
\includegraphics[width=0.45\textwidth]{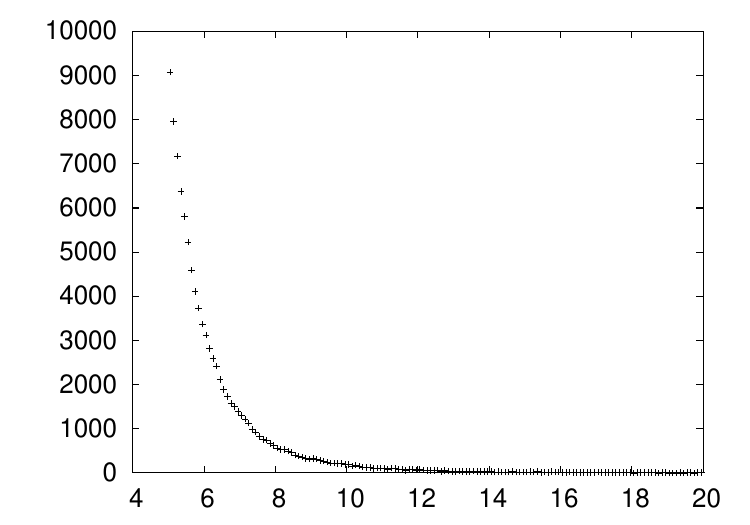}
\end{center}
\caption{Output of Pythia Monte Carlo simulation. Left plot: high energy neutron background (lower, St. Andrew's crosses) against a would-be gluon signal (upper, straight crosses) injected at the milibarn level into the simulation.
Right: run with increased statistics to show the background alone in somewhat more detail.
 \label{fig:neutronbackground}
}
\end{figure}

In figure~\ref{fig:neutronbackground} we present the result of the simulation, plotting number of particles produced as function of their total energy.

To address the background we generate large samples of events and look for those with an energetic neutron (plus any other particles that we discard). We impose a kinematic cut on rapidity  
\be
|\eta|  = \left|\tanh^{-1} \frac{p_z}{E} \right| <1
\ee %
that approximately matches Alice's detection capabilities.
We also demand that the neutron (or later, would-be gluon) produced has an energy of at least 5 GeV. The simulation is semiinclusive, any neutrons produced satisfying these requisites will be counted, irrespective of the rest of the event.

First we turn to the left plot of figure~\ref{fig:neutronbackground};
this results of a simulation in which 
we injected, in addition to the Standard Model particles,
a boson $\tilde{g}$ with spin $s=1$ and mass $m_0\approx 800\,\rm{MeV}$, that exits the collision without hadronizing (PYTHIA can be instructed to do this by not assigning color to the boson in the program~\cite{Boos:2001cv}, which is no problem for us since we control it analytically for other purposes).
We instruct the code to produce this boson in both
$gg\rightarrow\tilde{g}\tilde{g}$ and $q\bar{q}\rightarrow\tilde{g}\tilde{g}$ 
processes, with an isotropic diferential cross section given, in mbarn, by
\be 
\sigma = \frac{C}{p_T^2}\
\ee %
for each process, where $C$ is an arbitrary constant that we use to control the intensity of the signal.
The production cross sections added into PYTHIA that these $C$ factors control are, respectively,
$\sigma \approx 1.15\,\rm{b}$ and $\sigma \approx
0.15 \, \rm{b}$ for the $gg\rightarrow\tilde{g}\tilde{g}$ and
$q\bar{q}\rightarrow\tilde{g}\tilde{g}$ processes (comparable with the total cross sections which PYTHIA assigns to $gg\rightarrow gg$ and $gg\rightarrow q\bar{q}$ 
processes, which are $\sigma \approx 12.83\,\rm{b}$ and $\sigma \approx 0.09\,\rm{b}$,
respectively).
The total number of collisions employed for this plot was $10^8$.
The would-be gluon cross section production is at the 2 mbarn level (corresponding to some 89000 kept events), with the neutron background about a fourth as much, at 20000 events and 0.47 mbarn. 

In the right plot we take a closer look at the background alone with a higher number of collisions, reaching almost $5\times 10^8$, and confirm the half milibarn background for neutrons above 5 GeV. (If we demand the stricter 12 GeV criterion that we need for the application of Regge theory, this falls to the microbarn level, but it is more challenging to simulate properly).

It is clear from this exercise that, from the total number of counts alone, one cannot establish a bound better than 0.1 mbarn on the gluon liberation cross section in the $pp$ reaction in Eq.~(\ref{production}) (for energies down to 5 GeV). 

But one can do better. A little detector tomography, detecting the point of emission of secondary protons, allows excluding the would-be signal of liberated gluons
because of their much faster exponential fall-off. 

\begin{figure}
\begin{center}
\includegraphics[width=0.6\textwidth]{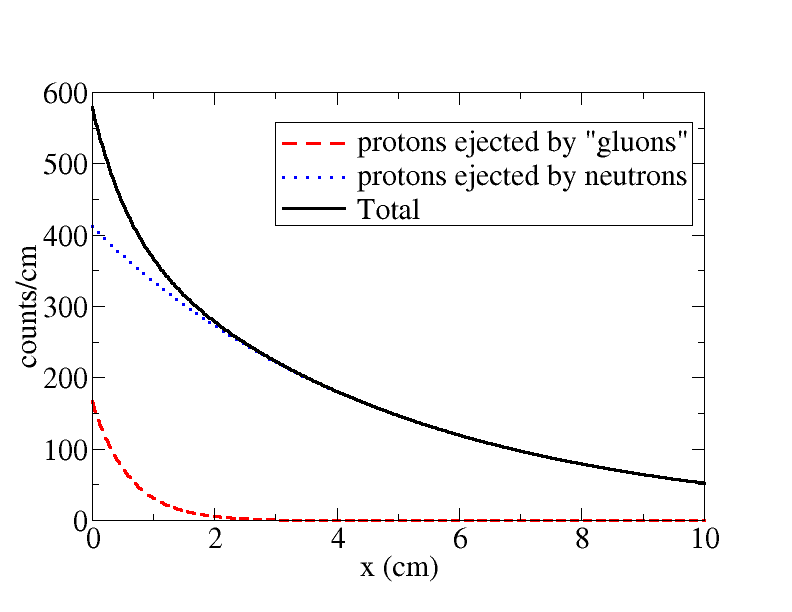}
\end{center}
\caption{Fall-off for proton production in the inner detector due to the nuetron background and the would-be gluon signal, at levels of 0.5 mbarn and 2$\mu$barn respectively. The much faster exponential fall of the gluon-induced emission (with shorter mean free path  in matter) allows to pick up a smaller signal.
 \label{fig:tomography}
}
\end{figure}

In figure~\ref{fig:tomography} we plot the number of ejected protons for an unspecified integrated luminosity, given an example 0.5 mbarn energetic neutron production cross-section, and 250 times less liberated gluon production, at the 2$\mu$barn level. Because the would-be gluons have a much shorted mean free path, they are captured earlier-on in the detector and provide a deviation from the simple exponential that characterizes the protons emitted by the neutron background.

Again, with the stricter cut at 12 GeV one could push this to the nbarn range.

\subsection{Further technical details}

We have used the random-number generator included in PYTHIA~\cite{Marsaglia}.
According to PYTHIA's documentation~\cite{sjostrand}, the intrinsic random number generator provides uniquely different random number sequences 
as long as the integer seeds remain below 900,000,000. 

The neutron background calculation reported in fig.~\ref{fig:neutronbackground}, left plot, consisted of a batch of 1000 simulations of $10^5$ events each,
with seeds varying from $1$ to $1000$ in each of the simulations. 

The gluon signal in the same plot was ran under identical conditions but with seeds varying from $2$ to $1001$. 

Finally, the right plot of fig.~\ref{fig:neutronbackground}
consists of 49 simulations of $10^7$ events, with seeds from $1$ to $50$ 
excluding $46$.

  The computation of all cross sections (and their errors, not shown but much smaller) is handily obtained by computing the mean value of the /sigmaGen()/ evaluations and the mean squared root of the /sigmaErr()/ evaluations obtained for each seed (see ref.~\cite{sjostrand}).

All simulations were ran at a standard linux-based cluster with 76 processors at the Universidad Complutense. The precise version of PYTHIA that we employed was 8.165, released on May 8th, 2012.

\section{Estimate of the total gluon-proton cross section}
\label{sec:sigmasec}

In this section we will support our statement that all potentially produced gluons would be slowed upon interacting in the inner parts of the experiment, by showing that the cross-sections $\sigma_{gp}$ will turn out to be large.

No experiment can directly access this secondary cross-section by preparing a beam of gluons, but we can estimate it theoretically employing the concept of parton-nucleon scattering amplitude~\cite{Landshoff:1970ff}.
The conjecture is motivated by the fact that quarks and gluons carry the strong force, and states that if a colored parton (gluon or quark) could ever be produced in isolation, it would undergo Regge scattering off a hadron just like any other hadron pair. This idea has been employed to derive Regge behavior of Deep Inelastic Scattering at low $x$~\cite{Brodsky:1973hm}, and recently employed to discuss the limitations of the Generalized Parton Distribution factorization theorem in exclusive processes~\cite{Szczepaniak:2007af}.

{
Regge theory parametrizes high-energy cross sections as simple powers of the energy. In the case of reactions where no quantum numbers are exchanged, the leading Regge pole is the so-called Pomeron
\be  \label{pomeroneq}
\sigma_{gp}(s) = \sigma'_{gp}(s) e^{b t}
\ee
with
\be
\sigma'_{gp}(s) \equiv \frac{4\pi^{2}}{\lambda^{\frac{1}{2}}(s,m_{g}^{2},m_{p}^{2})}\times f_{A}^{R}(t) f_{B}^{R}(t) \left(\frac{s}{\hat{s}} \right)^{\alpha(t)}
\ee
\be
\lambda(a,b,c) = a^{2}+b^{2}+c^{2}-2ab-2ac-2bc
\ee
with the Pomeron's Regge exponent $\alpha(0)\simeq 1$ yielding an approximately $s$-independent cross-section for 
$$
s_{gp}\in((5{\rm GeV})^2,(50{\rm GeV})^2)\ .
$$
}
Regge exponents are universal, in the sense that they depend only on the quantum numbers exchanged, and not on the precise hadrons that exchange them. Thus we expect the gluon-proton interaction to have the same Pomeron-dominated behavior of the total proton-proton cross-section.

Therefore we will estimate the secondary cross section to be proportional to the proton-proton reaction $\sigma_{pp}$ (well studied).
To determine the proportionality constant we will notice that the non-universal part of the Regge exchange is in the $\beta$ factors (in Feynman language, the vertex couplings of the Regge pole to the external hadron lines).
These couplings have been empirically determined for various physical meson-meson, baryon-baryon, meson-baryon, and photon-hadron reactions. But we need to estimate them theoretically in the case of gluon-hadron interactions.

To extrapolate from known physics, we recall that the pomeron in Quantum Chromodynamics is supposed to arise from a tower of glueball exchanges~\cite{Simonov:1990uq,LlanesEstrada:2000jw} with positive parity and charge conjugation. At least in the 
BFKL~\cite{Reisert:2009qv} domain at low Bjorken-$x$ one can derive Regge behavior from correlated two-gluon exchange in QCD.
Adopting this Pomeron-glueball connection, the differences between the pomeron coupling to various hadrons can be understood from the coupling of two gluons. 

Since the parton-nucleon scattering amplitude is a colored amplitude, its counting with the number of colors $N_c$ is much enhanced respect to color singlet-color singlet interactions. Therefore we are going to bypass momentum-space or spin wavefunction suppressions of order 1 and concentrate on these large color factors.

\subsection{Color factors for hadron-hadron scattering}\label{subsec:color}
As a warm-up, let's reproduce the color factors appropriate for the Pomeron-mediated pion-pion amplitude leading to the $\sigma_{\pi\pi}$  total cross-section.

\begin{figure}[H]
\includegraphics[width=0.15\textwidth]{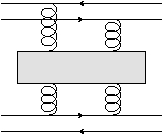}
\includegraphics[width=0.15\textwidth]{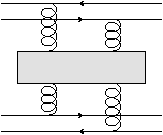}
\includegraphics[width=0.15\textwidth]{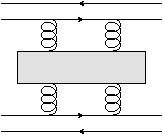}
\caption{Pomeron-mediated meson-meson interaction. \label{fig:meson}}
\end{figure}

Consider first the Feynman diagrams in figure~\ref{fig:meson} corresponding to two-gluon exchange between  two mesons.
\\
We employ the standard t'Hooft $N_c$ scaling in which $g(N_c)=g(3)\sqrt{3/N_c}$, and we remain at $N_c=3$ denoting $g(3)\equiv g$.
{The meson wavefunction is normalized by $1/\sqrt{N_c}$.
All three diagrams in the figure carry the same color factor equal to
\begin{eqnarray}
\nonumber{}
C_{MM} = \left( \frac{1}{\sqrt{N_{c}}}\right)^{4} \left( \frac{g}{\sqrt{N_{c}}} \right)^{4} T_{ij}^{(a)}T_{kl}^{(a)}T_{lk}^{(b)}T_{ji}^{(b)} =\\
\nonumber{}
\frac{g^{4}}{N_{c}^{4}} \frac{1}{2}\left(\delta_{mq}\delta_{lp}-\frac{1}{N_{c}}\delta_{lm}\delta_{pq} \right)\frac{1}{2}\left(\delta_{qm}\delta_{pl}-\frac{1}{N_{c}}\delta_{pq}\delta_{ml} \right) 
\ea
from where
\be
C_{MM}= \frac{1}{4} \frac{g^{4}}{N_{c}^{4}} \cdot (N_{c}^{2} - 1)
\ee
Substituting $N_{c} = 3$ we obtain
\begin{equation}
C_{MM} = \frac{2}{81} g^{4}
\end{equation}
}

Turning now to the baryon-baryon 
Pomeron-mediated interaction, there are three possible Feynman diagrams (in the lowest order of the vertex couplings) with different color factor. We consider them separately. Observe first the diagram in figure \ref{fig:baryon1}
\begin{figure}[H]
\begin{center}
\includegraphics[width=0.15\textwidth]{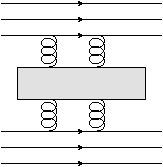}
\end{center}
\caption{Baryon-baryon Pomeron exchange with double quark-two gluon coupling
\label{fig:baryon1}.}
\end{figure}
Since baryons have at least $N_c$ quark constituents, the normalization is now $1/\sqrt{N_c!}$. This yields
\begin{eqnarray} \label{BB1}
C_{BB1}  = \\ \nonumber
 \left( \frac{g}{\sqrt{N_{c}}} \right)^{4} T_{kl}^{(a)}T_{pq}^{(a)}T_{qr}^{(b)}T_{lm}^{(b)} \frac{\epsilon_{ijk}}{\sqrt{N_{c}!}}\frac{\epsilon_{ijm}}{\sqrt{N_{c}!}}\frac{\epsilon_{pon}}{\sqrt{N_{c}!}}\frac{\epsilon_{ron}}{\sqrt{N_{c}!}}  \\
\nonumber
 = \frac{1}{\left( N_{c}! \right)^{2}}\left( \frac{g}{\sqrt{N_{c}}} \right)^{4} T_{kl}^{(a)}T_{pq}^{(a)}T_{qr}^{(b)}T_{lm}^{(b)}   \epsilon_{ijk}\epsilon_{ijm}  \epsilon_{pon}\epsilon_{ron} 
\\ \nonumber
=  \frac{1}{\left( N_{c}! \right)^{2}} \frac{g^{4}}{N_{c}^{2}} \cdot (N_{c}^{2} - 1)= \frac{4}{81} g^{4}
\mbox{} \nonumber
\end{eqnarray}
(for $N_c=3$).

We now turn to the second possible diagram represented in figure~\ref{fig:Baryon2}
\begin{figure}[H]
\begin{center}
\includegraphics[width=0.15\textwidth]{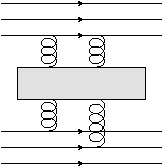}
\end{center}
\caption{Baryon-baryon interaction with an exchanged Pomeron and a single quark-two gluon coupling \label{fig:Baryon2}.}
\end{figure}
We obtain straightforwardly
\begin{eqnarray}
 \label{BB2}
C_{BB2}   = \\ \nonumber
\left( \frac{g}{\sqrt{N_{c}}} \right)^{4} T_{pq}^{(a)}T_{il}^{(a)}T_{qr}^{(b)}T_{jm}^{(b)} \frac{\epsilon_{ijk}}{\sqrt{N_{c}!}}\frac{\epsilon_{lmk}}{\sqrt{N_{c}!}}\frac{\epsilon_{onp}}{\sqrt{N_{c}!}}\frac{\epsilon_{onr}}{\sqrt{N_{c}!}} \\
\nonumber{}
 = \frac{1}{\left( N_{c}! \right)^{2}}\left( \frac{g}{\sqrt{N_{c}}} \right)^{4} T_{pq}^{(a)}T_{il}^{(a)}T_{qr}^{(b)}T_{jm}^{(b)}  \epsilon_{ijk}\epsilon_{lmk} \epsilon_{onp}\epsilon_{onr} 
\\ \nonumber
=  \frac{-1}{2 \left( N_{c}! \right)^{2}} \frac{g^{4}}{N_{c}^{2}} \cdot (N_{c}^{2} - 1)= \frac{-2}{81} g^{4}
\mbox{} \nonumber
\end{eqnarray}
And finally we have the diagram in figure~\ref{fig:Baryon3}
\begin{figure}[H]
\begin{center}
\includegraphics[width=0.15\textwidth]{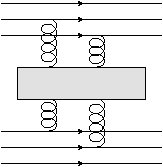}
\end{center}
\caption{Feynman diagram for baryon-baryon interaction with the two-gluons coupling to different quarks.
\label{fig:Baryon3}}
\end{figure}
that leads to
\begin{eqnarray} \label{BB3}
C_{BB3} =  \\ \nonumber
\left( \frac{g}{\sqrt{N_{c}}} \right)^{4} T_{jl}^{(a)}T_{pr}^{(a)}T_{km}^{(b)}T_{oq}^{(b)} \frac{\epsilon_{ijk}}{\sqrt{N_{c}!}}\frac{\epsilon_{ilm}}{\sqrt{N_{c}!}}\frac{\epsilon_{nop}}{\sqrt{N_{c}!}}\frac{\epsilon_{nqr}}{\sqrt{N_{c}!}} \\
\nonumber{}
= \frac{1}{\left( N_{c}! \right)^{2}}\left( \frac{g}{\sqrt{N_{c}}} \right)^{4} T_{jl}^{(a)}T_{pr}^{(a)}T_{km}^{(b)}T_{oq}^{(b)}   \epsilon_{ijk}\epsilon_{ilm} \epsilon_{nop}\epsilon_{nqr} 
\\  
\nonumber{}
=  \frac{1}{4 \left( N_{c}! \right)^{2}} \frac{g^{4}}{N_{c}^{2}} \cdot (N_{c}^{2} - 1)= \frac{1}{81} g^{4}
\mbox{} \nonumber
\end{eqnarray}

Comparing the color factors of the pion-pion and proton-proton amplitudes, of  the same order of magnitude for $N_c=3$, one would expect that the amplitudes and, consequently, the total cross sections, be also comparable.
 Experimental data~\cite{Nakamura:2010zzi} bear this expectation (compare the proton-proton cross-section $\sigma_{pp}(20~GeV) \simeq 40 -50$ mbarn with the $\sigma_{\pi^{+}\pi^{-}}(20~GeV) = 13.4 \pm 0.6$ mbarn $\pi\pi$ cross-section obtained in~\cite{Pelaez:2003ky,Pelaez:2004ab}).

\subsection{Proton and deconfined-gluon scattering}

Having reviewed the traditional cases of meson-meson and baryon-baryon scattering, our next step will be to calculate the gluon-baryon color factor and scale the high energy Regge parametrization of experimental  proton-proton scattering by this new color factor instead of that naturally corresponding to proton-proton.
The gluon carries the index of the adjoint representation of $SU(N_c)$
and the proton color analysis proceeds as in subsection \ref{subsec:color}

There are four possible couplings of the two Pomeron-generating gluons to the gluon-proton system. 
The first two, employing the three-gluon vertex, are depicted in figure~\ref{fig:gluon1}
\begin{figure}[H]
\begin{center}
\includegraphics[width=0.15\textwidth]{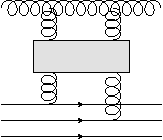}
\hspace{0.5cm}
\includegraphics[width=0.15\textwidth]{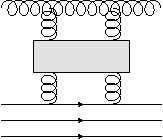}
\caption{Gluon-baryon interaction with an exchanged Pomeron featuring the  three-gluon vertex.\label{fig:gluon1}}
\end{center}
\end{figure}
These two diagrams carry as color factors
\begin{eqnarray}\label{gB1}
C_{gB1}  = \\ \nonumber \left(i \frac{g}{\sqrt{3}} f_{abd} \right) \left(i \frac{g}{\sqrt{3}} T^{(d)}_{km} \right) \left(i \frac{g}{\sqrt{3}} f_{bce} \right)  \left(i \frac{g}{\sqrt{3}} T^{(e)}_{jl} \right)  \\
\nonumber{}
\frac{\epsilon_{ijk}}{\sqrt{6}} \frac{\epsilon_{ilm}}{\sqrt{6}}= \frac{1}{54}g^{4}f_{abd}f_{bce}T^{(d)}_{km}T^{(e)}_{jl}\epsilon_{ijk}\epsilon_{ilm} = \\
\nonumber{}
=  \frac{-1}{54}g^{4}f_{abd}f_{bce}T^{(d)}_{km}T^{(e)}_{mk} \\ \nonumber
= \frac{1}{9} g^{4} \delta^{a}_{c}
\mbox{}
\end{eqnarray}
and
\begin{eqnarray}\label{gB2}
C_{gB2}  = \\ \nonumber \left(i \frac{g}{\sqrt{3}} f_{abd} \right) \left(i \frac{g}{\sqrt{3}} T^{(d)}_{kl} \right) \left(i \frac{g}{\sqrt{3}} f_{bce} \right)  \left(i \frac{g}{\sqrt{3}} T^{(e)}_{lm} \right)  \\
\nonumber{}
\frac{\epsilon_{ijk}}{\sqrt{6}} \frac{\epsilon_{ijm}}{\sqrt{6}} = \frac{1}{54}g^{4}f_{abd}f_{bce}T^{(d)}_{kl}T^{(e)}_{lm}\epsilon_{ijk}\epsilon_{ijm}  = \\
 = \frac{2}{54}g^{4}f_{abd}f_{bce}T^{(d)}_{km}T^{(e)}_{mk} \\ \nonumber
= \frac{-2}{9} g^{4} \delta^{a}_{c}
\mbox{} \nonumber
\end{eqnarray}
respectively.
The alternative two diagrams feature the four-gluon vertex and are depicted in figure~\ref{fig:gluon2}.
\begin{figure}[H]
\begin{center}
\includegraphics[width=0.15\textwidth]{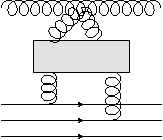}
\hspace{0.5cm}
\includegraphics[width=0.15\textwidth]{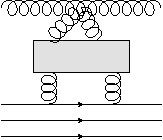}
\caption{Gluon-baryon interaction with an exchanged Pomeron featuring the  four-gluon vertex.\label{fig:gluon2}}

\end{center}
\end{figure}
The two color factors can easily be seen to vanish,\\
\begin{eqnarray}
\nonumber{}
C_{gB3}  = \\ \nonumber
 \left(- i \left(\frac{g^{4}}{\sqrt{3}}\right)^{2} [f_{abe}f_{cde}+f_{ade}f_{bce}+f_{ace}f_{dbe}]\right) \\
\nonumber{}
~~~~ \left(-i \frac{g}{\sqrt{3}} T^{(c)}_{km}\right) \left(-i \frac{g}{\sqrt{3}}T^{(d)}_{jl}\right) \frac{\epsilon_{ijk}}{\sqrt{6}} \frac{\epsilon_{ilm}}{\sqrt{6}} = \\
\nonumber{}
=  i \frac{g^{4}}{54}[f_{abe}f_{cde}+f_{ade}f_{bce}+f_{ace}f_{dbe}] \\
\nonumber{}
~~~~ T^{(c)}_{km}T^{(d)}_{jl} \left(\delta_{jl}\delta_{km} - \delta_{jm}\delta_{kl} \right)= \\
\nonumber{}
= - i \frac{g^{4}}{54}[f_{abe}f_{cde}+f_{ade}f_{bce}+f_{ace}f_{dbe}] T^{(c)}_{km}T^{(d)}_{mk} = \\
\nonumber{}
= - i\frac{g^{4}}{27} [f_{abe}f_{cce}+f_{ace}f_{bce}+f_{ace}f_{cbe}]  
\\ \nonumber =0
\end{eqnarray}
and
\begin{eqnarray}
\nonumber{}
C_{gB4} = \\ \nonumber
\left(- i \left(\frac{g^{4}}{\sqrt{3}}\right)^{2} [f_{abe}f_{cde}+f_{ade}f_{bce}+f_{ace}f_{dbe}]\right) \\
\nonumber{ }
~~~~ \left(-i \frac{g}{\sqrt{3}} T^{(c)}_{kl}\right) \left(-i \frac{g}{\sqrt{3}}T^{(d)}_{lm}\right) \frac{\epsilon_{ijk}}{\sqrt{6}} \frac{\epsilon_{ijm}}{\sqrt{6}} = \\
\nonumber{}
= -  i \frac{g^{4}}{54}[f_{abe}f_{cde}+f_{ade}f_{bce}+f_{ace}f_{dbe}] \\
\nonumber{}
~~~~ T^{(c)}_{kl}T^{(d)}_{lm} 2 \delta_{km}= \\
\nonumber{}
=  2i \frac{g^{4}}{27}[f_{abe}f_{cde}+f_{ade}f_{bce}+f_{ace}f_{dbe}] T^{(c)}_{kl}T^{(d)}_{lk} = \\
\nonumber{}
= 2 i\frac{g^{4}}{27} [f_{abe}f_{cce}+f_{ace}f_{bce}+f_{ace}f_{cbe}]  
\\ \nonumber =0
\end{eqnarray}
because two of the gluons out of the four-gluon vertex are forced to exit in a color singlet (as appropriate for color-singlet Pomeron exchange).

Once the color factors have been computed, we just divide the  proton-proton amplitude by the typical $1/81$ from Eq.~(\ref{BB1}), (\ref{BB2}) and (\ref{BB3}), and multiply it by the $1/9$ from 
Eq.~(\ref{gB1}) and (\ref{gB2}).

Squaring, we expect the gluon-proton cross-section to be a factor of 81 times the proton-proton cross-section. 
In figure \ref{fig:fit} we plot the total proton-proton cross-section 
and its color rescaling to give what we theorize to be the total gluon-proton cross-section.

After including the lower limit for the momentum transfer to the secondary proton in ALICE (due to the TOF identification requirement), and following~\cite{Doyle:1996dk} for the $b$-exponential slope, the possible gluon-proton cross section is depicted in figure~\ref{fig:fit2}.

\begin{figure}
\begin{center}
\includegraphics[width=0.50\textwidth]{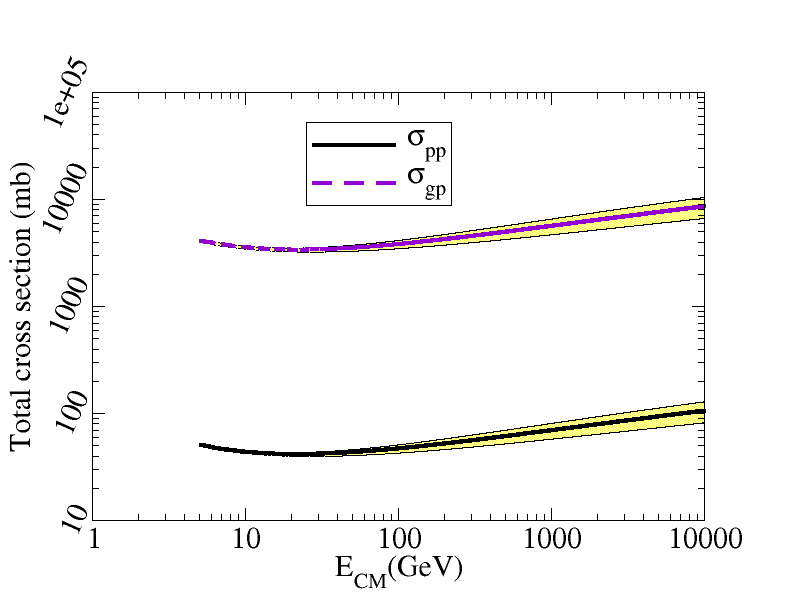}
\end{center}
\caption{ $\sigma_{pp}$ (bottom) and $\sigma_{gp}$ (larger at the top) as function of the secondary collision's center of mass energy $\sqrt{s_{gp}}$. \label{fig:fit}}
\end{figure}

\begin{figure}
\begin{center}
\includegraphics[width=0.50\textwidth]{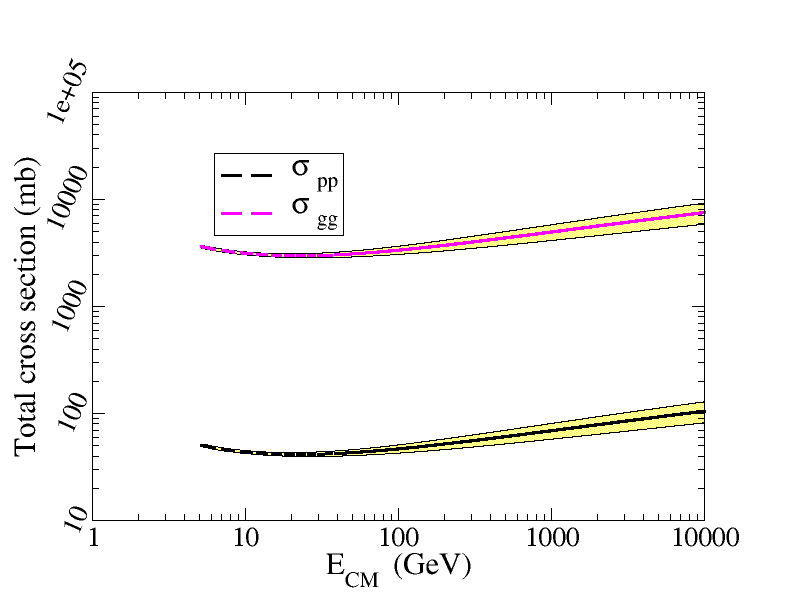}
\end{center}
\caption{ $\sigma_{pp}$ (bottom) and $\sigma_{gp}$ (top) as in figure~\ref{fig:fit} but with a minimum cut in the $\ar t\ar $ transfer to the secondary proton in the $gp$ interaction to insure a minimum velocity for it to reach the time of flight detector. \label{fig:fit2}}
\end{figure}

\newpage
\section{Discussion}

\footnote{Much of this is based on exchanges with anonymous referees}The gluon-proton cross section that we have estimated and can be seen in figure~\ref{fig:fit2} might also be of interest for mixed hadron-quark and gluon phases that could exist in heavy-ion collisions. Thermodynamics (for example, through the virial coefficients that depend on the interaction) as well as transport (the scattering rate enters the Boltzmann equation) calculations are bound to require estimates of gluon-hadron scattering. There is even a planned facility in whose physics programme looking for this mixed phase occupies a very prominent place~\cite{Musulmanbekov:2011zz}, and other existing or planned experiments can probably also look for signatures of this mixed phase (ALICE at the LHC, STAR at RHIC, or CBM at FAIR).

Concerning the propagating properties of the deconfined gluon, we wish to note that if the gluon dispersion relation is that of a massless particle, the (dominant) $\pi_0$ decay width to two photons strongly establishes gluon confinement. 
This is because of the good agreement between experiment and theory at the few percent level in estimating this width; since the quark-gluon color coupling is so much stronger than the electric quark-photon coupling, a pion decaying to two massless gluons would have a larger than observed hadronic width.

However the absence of strong $\pi_0$ decays can also be hindered by a non-trivial dispersion relation. In addition to numerous lattice and Dyson-Schwinger evidence, the largely discrete spectrum itself shows that gluons are gapped. Otherwise there would be a continuum of states with all quantum numbers below ordinary thresholds down to the pion mass. 

Thus, is the absence of a strong pion width suppressed because of confinement, or because of kinematic hindrance? If we take the lattice glueball mass of 1.5-1.7 GeV (surely a gauge-invariant quantity) as indicative of the relevant gap scale, no narrow meson has enough phase space to decay to two gluons.

Gluons as field quanta are gauge-dependent entities. We give them physical meaning as parcels of energy and momentum emitted in a single impulse and revealed through a hadron jet. The question at hand is whether these parcels can also continue propagating as one neutral particle. The answer from $\pi_0$ decays is: not if they are massless. An interesting cartoon of the various possibilities is figure III.I of reference~\cite{Bashir:2012fs}. One can conceivably have screening without confinement, that is, a gapped spectrum without color trapping. They are independent aspects requiring separate testing. 

$\Upsilon$ mesons on the other hand (page 1110 of~\cite{Nakamura:2010zzi}) have an estimated branching fraction to three gluons of order 81.7(7)\% . Of this fraction, $90\%$, that is, about 40(4) keV, are dark, unreconstructed decays. This is a bound to the decay width to three unconfined gluons. 
We can turn this bound in turn into a bound on the  $b\bar{b}\to ggg$ cross section for a $b$ quark and antiquark to yield three unconfined gluons. The estimate involves the definition of the cross section as number of reactions per unit time, divided by the number of crossing pairs per unit area and time, and it also requires the interpretation of the non-relativistic wavefunction square as a volume probability density. Since the number of reactions per unit time for a properly normalized bottomonium wavefunction is precisely the inverse width,
\be
\Gamma \sim \sigma v \ar \psi(0)\ar^2 \ . 
\ee
And more, since near the origin the potential between heavy quarks can be taken as Coulombic,
\be
\sigma \sim \frac{\Gamma}{m_b^3 \alpha_s(m_b)^4}\ .
\ee
That the decay branch to three deconfined gluons needs to be smaller than the as yet unreconstructed dark decays $\Gamma_{ggg}<40(4)$ keV leads therefore to the bound for the $b$ quark-antiquark pair to annihilate into three gluons with deconfined color
\be
\sigma_{b\bar{b}\to ggg} \le 100 {\rm nbarn} \ .
\ee
This bound should be trusted to within order of magnitude only; a similar analysis applied to orthopositronium and used to predict its three photon width given the QED cross-section for $e^-e^+\to \gamma\gamma\gamma$ misses by about a factor 3. It is obviously a very loose one and a quick measurement at a modern collider can 
potentially constrain deconfined gluon production in terms of asymptotic particles instead of quarks, with similar precision at the microbarn level and better, with an ultimate limit at the 10-femtobarn scale.

Once again, we stress that we do not take theoretically seriously the possibility of emitting a massive, deconfined gluon (vector constituent of hadrons). We just contend that it is not empirically closed.

To summarize, we have argued that gluon confinement is only weakly established experimentally. By assuming that a gluon might have been deconfined in a high energy experiment, we have asked ourselves what the characteristic signature would be. 

With minimum theoretical assumptions (Regge gluon-nucleon interaction) we think that the emission of a secondary proton from the inner detector material would be the tell-tale signal. The background to this reaction is the production of neutrons, but their color-singlet interaction is much smaller than the gluon's, allowing to set meaningful bounds on the cross section for free gluon production, that we estimate at the microbarn level. The ultimate goal conceivable with more sophisticated analysis is $O(10)$ fbarn for collider energies.

For comparison, one can think of the 1-50 pbarn typical of supersymmetric particle -notably gluino- searches.

\begin{acknowledgements}
Work supported by grants FPA 2008-00592, FIS2008-01323,
FPA2007-29115-E, PR34-1856-BSCH, UCM-BSCH \\ GR58/08 910309, PR34/07-15875 (Spain), and a Complutense-GSI exchange grant.
\end{acknowledgements}


\end{document}